\documentclass[reprint,amsmath,amssymb,aps,prl,superscriptaddress]{revtex4-1}

\usepackage{graphicx}
\usepackage{color}
\usepackage{amsmath, bm} 
\usepackage[version=3]{mhchem} 
\usepackage{graphicx}
\usepackage{physics}
\usepackage{dcolumn}
\usepackage{bm}
\usepackage{hyperref}
\usepackage{color}

\newcommand{\AlGaAs}{\ce{Al_{x}Ga_{1-x}As}}

\newcommand{\code}[1]{\texttt{\detokenize{#1}}}
  
\begin{document}

\title{Anharmonic Lattice Relaxation during Non-radiative Carrier Capture}

\author{Sunghyun Kim}
\email{sunghyun.kim@imperial.ac.uk}
\affiliation{Department of Materials, Imperial College London, UK}
\author{Samantha N. Hood}
\affiliation{Department of Materials, Imperial College London, UK}
\author{Aron Walsh}
\email{a.walsh@imperial.ac.uk}
\affiliation{Department of Materials, Imperial College London, UK}
\affiliation{Department of Materials Science and Engineering, Yonsei University, Seoul 03722, Korea}

\date{\today}

\begin{abstract}
Lattice vibrations of point defects are essential for understanding non-radiative electron and hole capture in semiconductors as they govern properties including persistent photoconductivity and Shockley-Read-Hall recombination rate.
Although the harmonic approximation is sufficient to describe a defect with small lattice relaxation,
for cases of large lattice relaxation it is likely to break down.
We describe a first-principles procedure to account for anharmonic carrier capture and apply it to the important case of the \textit{DX} center in GaAs.
This is a system where the harmonic approximation grossly fails.
Our treatment of the anharmonic Morse-like potentials accurately describes the observed electron capture barrier, predicting the absence of quantum tunneling at low temperature, and a high hole capture rate that is independent of temperature.
The model also explains the origin of the composition-invariant electron emission barrier.
These results highlight an important shortcoming of the standard approach for describing point defect ionization that is accompanied by large lattice relaxation,
where charge transfer occurs far from the equilibrium configuration.
\end{abstract}

\maketitle


Following the pioneering work of Landau and Zener \cite{Zener:1932iz}, non-radiative charge transfer has been studied extensively in molecules and biological systems \cite{Marcus:1985ev}, as well as in condensed matter \cite{Huang:1950fg,Henry:1977bfa}.
In the Landau-Zener formula,
the probability of charge transfer is proportional to the square of the coupling of initial and final states
and inversely proportional to the rate of change in their energy spacing.
For point defects in crystalline solids, with quantized vibrations \cite{Huang:1950fg,Stoneham:gb,Henry:1977bfa, Shi:2012iv, Alkauskas:2014kk}, 
the carrier capture coefficient $C$ can be expressed 
using the electron-phonon coupling $\mel{\psi_i}{\pdv*{H}{Q}}{\psi_f}$
and the overlap of vibrational wave functions $\mel{\xi_{im}}{\Delta Q}{\xi_{fn}}$,
which is given by 
\begin{equation}
\begin{split}
    C = & \frac{2\pi}{\hbar} \abs{\mel{\psi_t}{\pdv*{H}{Q}}{\psi_c}}^2 \\
        & \sum_{m,n} w_{m} \abs{\mel{\xi_{tn}}{\Delta Q}{\xi_{cm}}}^2
    \delta(E_{cm} - E_{tn})
\end{split}
\end{equation}
where $\psi$ and $\xi$ are electronic and vibrational wave functions, respectively, and 
the subscripts $c$ and $t$ specify the free carrier and trap states.
$E_{cm}$ and $E_{tn}$ denote the energy of the carrier and trap states, respectively,
where $m$ and $n$ are the indices for vibrational eigenstates.
Here, we use the effective configuration coordinate $Q$.

In this formalism, the temperature-dependence is determined by the thermal occupation number $w_{m}$ of the initial vibrational state.
Early theories provided a good understanding of carrier capture rates that follow Arrhenius behaviour at high temperature and are limited by quantum mechanical tunneling at low temperature.
However, it is impossible to access the detailed parameters experimentally, including electron-phonon coupling matrix elements and overlap integrals of vibrational wave functions.
Instead, the weighted average of the capture rate is measured.

Modern simulation approaches have been based on density functional theory (DFT) that avoid the need for empirical parameters.
Shi and Wang \cite{Shi:2012iv} proposed an adiabatic formalism to calculate the capture rate using DFT, taking into account the full set of phonon modes for a given defect.
Later, a method adopting a 1D configuration coordinate and static coupling theory was proposed by Alkauskas \textit{et al.} \cite{Alkauskas:2014kk}.
The two methods have been compared, and the validity of the 1D configuration coordinate has been confirmed \cite{Shi:2015de,Wickramaratne:2018iv,Shi:2018rp}.
The adiabatic approximation generally underestimates the capture rate compared to measurements and static coupling theory \cite{Shi:2015de,Wickramaratne:2018iv,Shi:2018rp}.
However, as they both adopt the harmonic approximation for the potential energy surface (PES) of defects,
the role of anharmonicity in non-radiative carrier capture has not been well characterized.
The vibrations of a defect may differ significantly from harmonic behavior 
when its atomic configuration is far from the equilibrium structure, as has been suggested in earlier work \cite{Stoneham:gb,Henry:1977bfa,Markvart:1981ki}.

Carrier capture that occurs far from an equilibrium configuration has been understood
based on a large-lattice-relaxation (LLR) model developed by Lang and Logan \cite{Lang:1977ub}.
The LLR model successfully explained persistent photoconductivity (PPC) in \ce{Al_{x}Ga_{1-x}As} ($x>0.22$) and \ce{GaAs} under hydrostatic pressure
with the lack of quantum tunneling at low temperature ($< 77$ K) and a large Stokes shift of c.a. 1 eV.
The model adopts the harmonic approximation with empirical paramaterization; although, the authors mentioned that anharmonic terms may be important.

One of the most intensively studied LLR defects is the \textit{DX} center in \ce{Al_{x}Ga_{1-x}As},
owing not only to its anomalous physical properties but also its technological importance.
The isolated substitutional Si atom (\ce{Si_{Ga}}) was proposed as an atomic model of the \textit{DX} center by Chadi and Chang \cite{Chadi:1988eu, Chadi:1989el},
and successfully explained experimental observations.
However, a detailed microscopic understanding has not been fully explored.
Unusually, the electron emission barrier is invariant with respect to the variation in the composition of \AlGaAs~ and hence the donor binding energy \cite{Calleja:1988gk, Mooney:1990bx}.
With only circumstantial evidence, it has been assumed that the transition from the conduction band minimum at the $\Gamma$-point to \textit{DX} center is forbidden, and 
a hypothetical intermediate state, presumably related to the $\mathrm{L}$ valley of the conduction band,
plays an important role.

In this Letter, we report a first-principles anharmonic approach to describe non-radiative electron and hole capture in semiconductors.  
We apply it to investigate carrier trapping by the \textit{DX} center, \ce{Si_{Ga}}, in \ce{GaAs} under hydrostatic pressure.
During the atomic transformation accompanying carrier capture, the bond-breaking relaxation of \ce{Si_{Ga}} results in a Morse-like PES.
Here the harmonic approximation significantly overestimates the electron and hole capture barriers and fails to even qualitatively describe the physical behaviour of the system.

Our procedure is implemented in the open-source \code{CarrierCapture.jl} package \cite{CarrierCapture}.
We followed static coupling theory, but removed the restriction of harmonic vibrations.
Instead we calculate the vibrational wave functions $\xi$ and matrix elements $\mel{\xi_{tn}}{\Delta Q}{\xi_{cm}}$
from solutions of the 1D Schr\"{o}dinger equation for the anharmonic PES using a finite-difference method.
The total energy of pristine and defective crystals was calculated from DFT~\cite{Hohenberg1964, Kohn1965} using the projector-augmented wave (PAW) method \cite{Blochl1994} and the hybrid exchange-correlation functional of Heyd-Scuseria-Ernzerhof (HSE06) \cite{Heyd2003}, as implemented in \ce{VASP}~\cite{Kresse1999}.
We used a value of screened exact exchange ($\alpha=0.28$) that reproduces the experimental band gap of GaAs.
The wave functions were expanded in plane waves up to an energy cutoff of 400 eV.
The all-electron wave functions were derived from the pseudo wave functions and atom-centered partial waves in the PAW method,
and the overlap integrals were performed in real space using \code{pawpyseed} \cite{pawpyseed} to calculate the electron-phonon coupling outlined by Alkauskas \textit{et al.}~\cite{Alkauskas:2014kk}.
A Monkhorst-Pack $k$-mesh~\cite{Monkhorst1976} with a grid spacing less than $2\pi\times$0.03 \AA$^{-1}$ was used for Brillouin zone integration.
The atomic coordinates were optimized until the residual forces were less than 0.01 eV/\AA. %
The lattice vectors were relaxed until residual stress was below 0.5 kbar under external pressure of 28 kbar, which is a regime where the DX-centre is stable.
For defect formation, a $3\times3\times3$ supercell expansion (216 atoms) of the conventional cell was employed with $\Gamma$-point sampling to avoid spurious dispersion of the Kohn-Sham eigenstates for the defect. 

\begin{figure}
    \centering
    \includegraphics[width=7.0cm]{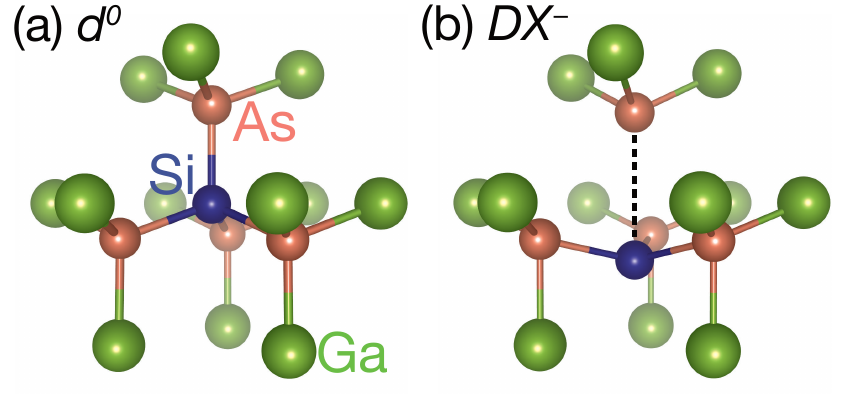}
    \caption{Atomic structures of \ce{Si_{Ga}} in (a) the neutral \textit{d} and (b) negatively-charged \textit{DX} configurations. 
    Only second-nearest-neighbor atoms are shown, for clarity. 
    For the \textit{DX} configuration, the broken Si--As bond is represented by a dashed line.
    }
    \label{fig:structure}
\end{figure}

\textit{Anharmonicity of the DX centre:}
\ce{Si_{Ga}} is a shallow donor in \ce{GaAs}.
The defect with \ce{T_{d}} symmetry is referred to as the \textit{d} configuration (Fig.~\ref{fig:structure} (a)).
In both \ce{GaAs} under hydrostatic pressure and the \ce{Al_{x}Ga_{1-x}As} ($x>0.22$) alloy, the band gap is widened and a shallow-to-deep transition occurs.
A deep donor, the so-called \textit{DX} configuration with \ce{C_{3v}} symmetry, becomes stable ((Fig.~\ref{fig:structure} (b)).
In the \textit{DX} configuration, the Si--As bond is broken and
the Si atom exhibits a large lattice relaxation toward the antibonding site.
We find that the neutral \textit{d} configuration produces a shallow level with a delocalized Kohn-Sham eigenstate, while
the negatively-charged \textit{DX} produces a deep eigenstate localized around a Si--As antibonding orbital.
This doubly occupied antibonding level stabilizes the \textit{DX} configuration.

\begin{figure}
    \centering
    \includegraphics[width=8.3cm]{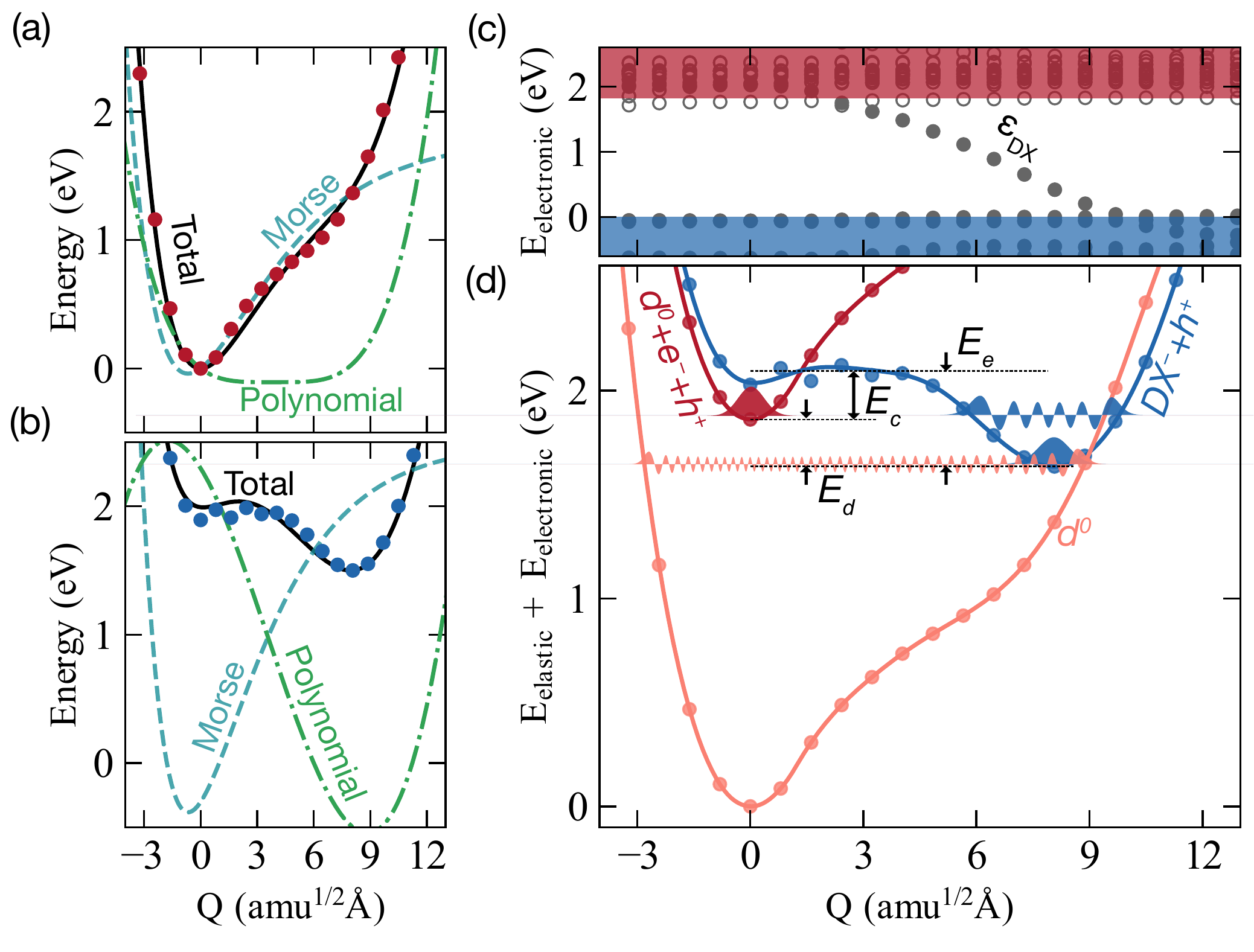}
    \caption{
    Potential energy surfaces of (a) \textit{d}$^0$ and (b) \textit{DX}$^-$ configurations of \ce{Si_{Ga}} in GaAs.
    The black solid line shows the best fit to the DFT data represented by solid circles.
    The best fit curves are composed of the Morse function (blue dashed line) and polynomial functions (green dash-dotted line).
    (c) Evolution of Kohn-Sham eigenstates of the supercell containing \textit{DX}$^-$ with respect to the deformation of the geometry along $Q$.
    Solid and empty circles represent occupied and unoccupied states, respectively.
    (d) Configuration coordinate diagram for electron and hole capture.
    The solid circle and lines depict the DFT results and the quadratic spline fit to the DFT data, respectively.
    The vibrational wave functions for the ground-state of initial PES and corresponding final state are shown.
    }
    \label{fig:cc}
\end{figure}

%
To begin, we mapped the PES of the \ce{Si_{Ga}} defect over a configuration coordinate $Q$ that represents the degree of collective atomic deformation (Fig.~\ref{fig:cc}).
$Q$ is defined by 
\begin{equation}
Q^{2} = \sum_{\alpha} M_{\alpha} \vb{R}_{\alpha}^{2}
\end{equation}
where $M_{\alpha}$ and $\vb{R}_{\alpha}$ are the atomic mass and the displacement vector from the equilibrium position of atom $\alpha$, respectively. 
The PES of \textit{d} and \textit{DX} are well described by a combination of two forms: a Morse potential and a polynomial potential up to fourth-order 
that describe the \ce{Si-As} bond-breaking, and additional bond-stretching and bond-bending around the defect, respectively.
For neutral \textit{d}$^0$, the stretching of the \ce{Si-As} bond increases the potential energy significantly over a short range,
but after the bond-breaking ($Q>4$ amu$^{1/2}$\AA), the energy increases moderately due to the bending of other bonds (Fig.~\ref{fig:cc} (a)).
Further distortion results in a substantial energy penalty.

For the charged \textit{DX$^-$}, the polynomial potential shows a minimum around the \textit{DX} configuration (8.8 amu$^{1/2}$\AA),
while the minimum of the Morse potential is near the \textit{d} configuration ($-0.7$ amu$^{1/2}$\AA).
Thus, the Morse potential describes the attractive force between Si and As
compensating the restoring force due to the perturbation of other bonds (Fig.~\ref{fig:structure} (b)).
This competition results in the soft anharmonic PES of \textit{DX}$^{-}$ near the \textit{d} configuration.
Further distortion results in strong Pauli repulsion which takes an exponential form.
Despite their simplicity, 
the combination of Morse and polynomial potentials adequately describe the DFT energy surface and produce a physical dissociation energy of 2--3 eV for the Morse component.

The evolution of the electronic eigenstate during the lattice relaxation from \textit{DX}$^-$ to \textit{d}$^0$ is shown in Fig.~\ref{fig:cc} (c).
As the Si and As atoms approach, the anti-bonding level rises toward the conduction band edge; at $Q = 2.4$ amu$^{1/2}$\AA ~ they cross.
It is challenging to describe diabatic level-crossing within the framework of DFT / Born-Oppenheimer approximation due to variational collapse.
Our practical solution is to employ a $\Delta$ self-consistent field  approach and constrain the occupation of the defect level. 
This approach recovers the diabatic process, but results in some noise near the crossing-point (Fig.~\ref{fig:cc} (b)) as the adiabatic basis is strongly coupled.
The development and application of more sophisticated excited-state techniques such as time-dependent DFT is a worthwhile line of research. 

The full configuration coordinate diagram describing electron capture (Fig.~\ref{fig:cc} (d)) is obtained
by aligning the PES of \textit{d}$^0$ and \textit{DX}$^-$ 
using the donor binding energy ($E_d$) of the \textit{DX} center,
which varies from 0--0.23 eV depending on AlAs mole fraction $x$ in \AlGaAs ~\cite{Mooney:1987bf} and the hydrostatic pressure \cite{Calleja:1988gk,Fujisawa:1988et,Fujisawa:1989ie}.
While the decomposition of the PES into intuitive functions is useful for qualitative analysis, we use quartic splines for the best fit to the DFT energy surface (Fig.~\ref{fig:cc} (d)). 

\begin{figure}
    \centering
    \includegraphics[width=8.3cm]{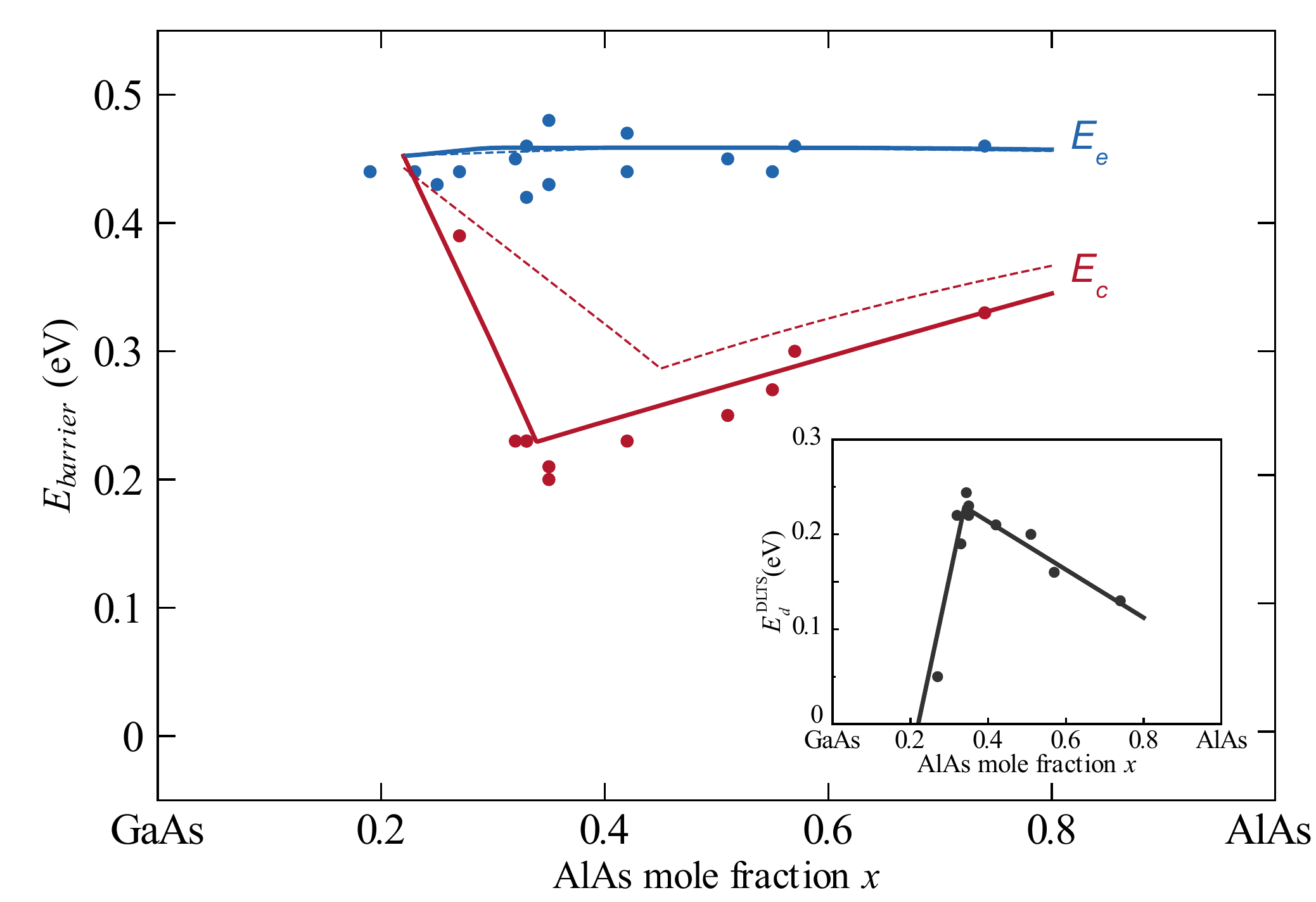}
    \caption{
    (a) Energy barriers for electron emission (blue line) and electron capture (red line) in \AlGaAs.
    The experimental data (filled circles) are taken from Ref.~\cite{Mooney:1987bf}.
    The calculated energy barriers are presented as a function of
    the molar fraction $x$ estimated using 
    the empirical donor binding energy ($E_{d}$) measured by DLTS \cite{Mooney:1987bf} (solid lines) and Hall experiments \cite{Chand:1984as} (dashed lines).
    The inset shows the donor binding energy $E_{d}^\mathrm{DLTS}$ (solid line) obtained from the best fit to the DLTS data (filled circles).
    }
    \label{fig:barrier}
\end{figure}

One anomaly of the \textit{DX} center is that 
the activation energy for electron emission $E_{e}$ is nearly invariant with respect to the \ce{AlAs} mole fraction $x$ and, hence, the donor binding energy as shown in Fig.~\ref{fig:barrier}.
This weak variation in $E_{e}$ can not be explained by configuration coordinates within the harmonic approximation.
We have calculated the emission barrier with various donor binding energies (Fig.~\ref{fig:barrier}).
The \ce{AlAs} mole fraction $x$ is estimated based on the donor binding energy measured by deep level transient spectroscopy (DLTS).
Due to the plateau in the potential of \textit{DX}$^{-}$ around $Q = 3$,
the activation energy is fixed to 0.45 eV above the vibrational ground-state, regardless of the donor binding energy.
Thus, it is the anharmonicity  of \textit{DX}$^{-}$ that results in the constant $E_{e}$.
The intermediate state that has been proposed previously \cite{Saxena:1982jg, Mooney:1990bx} is not required.
Furthermore, the calculated energy barrier for electron capture also agrees well with experiments \cite{Mooney:1987bf,Calleja:1988gk}.
The donor binding energy measured by Hall experiments (Fig.~\ref{fig:barrier})
results in slightly higher capture energy,
as they predict shallower levels \cite{Chand:1984as}.

\textit{Rate of carrier capture:}
The anharmonic PES significantly lowers the electron capture barrier as compared to predictions within the harmonic approximation. 
The calculated electron capture barriers are 0.2--0.4 eV, depending on the donor binding energy, 
which agrees well with experiments \cite{Mooney:1987bf}.
The harmonic approximation predicts much higher barriers of 1.5--1.6 eV.

\begin{figure}
    \centering
    \includegraphics[width=8.3cm]{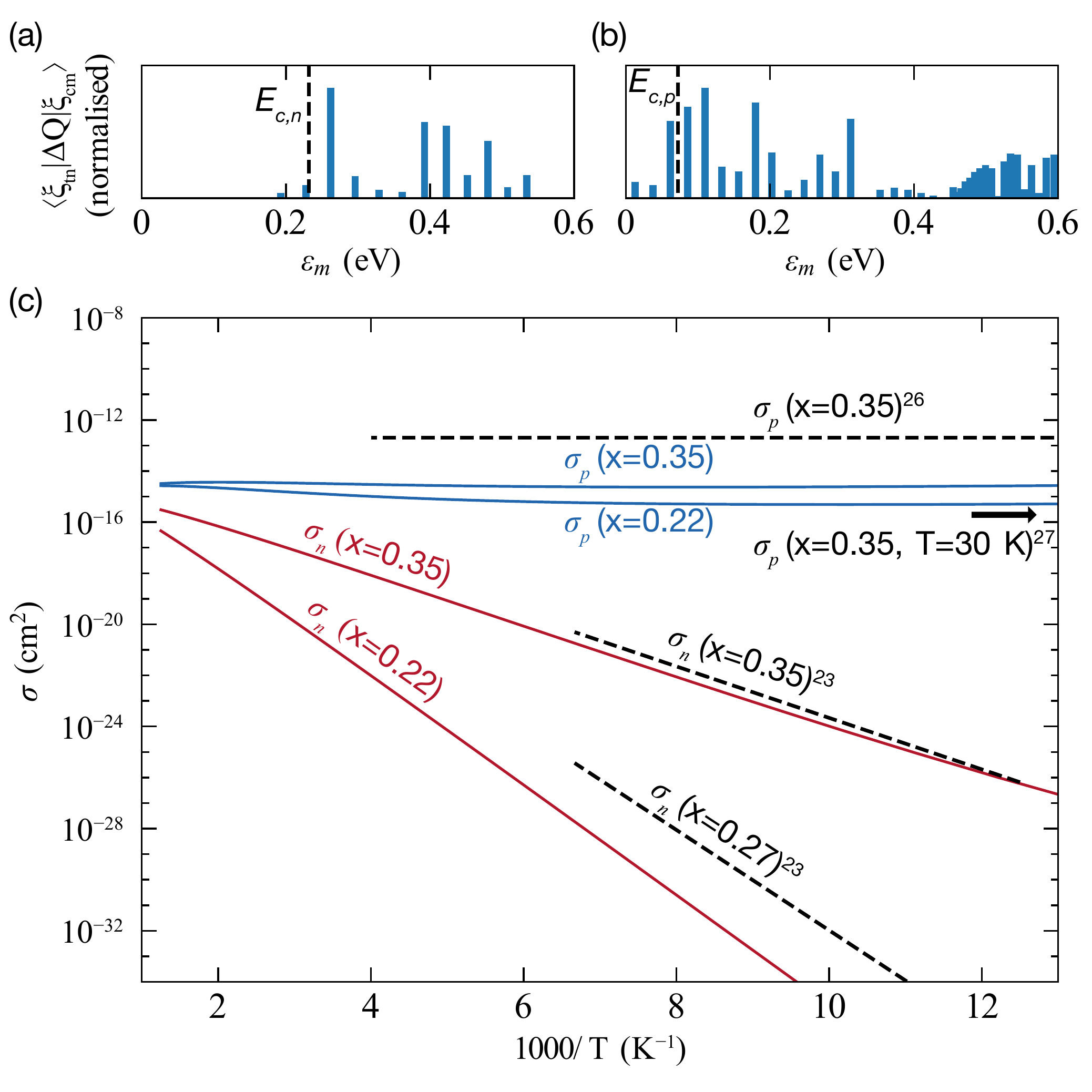}
    \caption{
    Overlap of vibrational wave functions with energy $\epsilon_m$
    for (a) electron capture and (b) hole capture of the \textit{DX} center with a donor binding energy of 0.2 eV. $E_{c,n}$ and $E_{c,p}$ represent the electron and hole barrier heights, respectively. 
    (c) Blue and red solid lines represent calculated electron and hole capture cross-sections of the \textit{DX} center. 
    The measured cross-sections (dashed lines and arrows) are taken from 
    Ref. \cite{Mooney:1987bf} for electron capture and Ref. \cite{Brunthaler:1989fs,Dobaczewski:1992hm} for hole capture.
    }
    \label{fig:capt}
\end{figure}

Next, we calculate the electron capture cross-section 
\begin{equation}
\sigma_{n} = C_{n} / \ev{v_{th}}
\end{equation}
where the thermal velocity $\ev{v_{th}}=\sqrt{3k_{B}T/m^{*}}$ is calculated using the effective masses $m^{*}$ of carriers taken from Ref. \cite{Levinshtein:2012fv}.
The high-temperature behavior of carrier capture is often governed by a classical energy barrier
while, at low temperature, tunneling is dominant. 
However, for the \textit{DX} center, 
the overlap of vibrational wave functions is negligible below the energy barrier, 
as shown in Fig.~\ref{fig:capt} (a),
due to large lattice relaxation and the long plateau in energy of \textit{DX}$^-$ (Fig.~\ref{fig:cc} (d)).
Thus, tunneling is suppressed and its effect is negligible. 
The calculated electron capture cross-section decreases exponentially, even at low temperature of around 77 K (Fig.~\ref{fig:capt} (c)), 
which explains the experimental observations \cite{Mooney:1987bf}.

In contrast, hole capture occurs with large overlap of vibrational wave functions even below the small energy barrier,
as shown in Fig.~\ref{fig:capt} (b).
This explains the weak temperature dependence in the hole capture cross-sections.
Moreover, the parallel PES ($Q > 9$ amu$^{1/2}$\AA) of \textit{d}$^{0}$ and \textit{DX}$^-$ 
produce large overlap populations above the crossing point (Fig.~\ref{fig:capt} (b)).
Here, the harmonic approximation predicts much smaller cross-sections due to the high hole capture barrier of 1 eV. 
On the other hand, the Morse component alone does not cross the potential energy surface of \textit{DX}$^{-}$,
which predicts that the vibrational wave function is unbounded, and only radiative recombination is allowed.
The restoring force on the Si atom provided by the remaining three \ce{Si-As} bonds 
ensures bound states with high energy and large hole capture cross-section.
After the hole capture, the excess energy is dissipated by emitting multiple phonons,
which is mainly attributed to the Morse potential, forming the \ce{Si-As} bond (Fig.~\ref{fig:cc} (a)).

In summary, we have shown that anharmonicity can play an important role in the non-radiative carrier capture process mediated by defects in semiconductors.
Due to the bond-breaking relaxation by the \textit{DX} center in GaAs, we find large lattice relaxation with a Morse-like potential.
The abnormal insensitivity of the activation energy for electron emission to \ce{AlAs} mole fraction can be explained by  anharmonicity of the potential.
The calculated carrier capture cross-sections of \textit{DX} center agree well with experiments
and differ significantly to those predicted from the harmonic approximation.
The anharmonic potential energy surfaces of the \textit{DX} center enhance the hole capture process and make it weakly dependent on temperature.
Thus we conclude that the harmonic approximation is insufficient when the charge transition occurs far from the equilibrium configuration, even if the full phonon spectrum is considered.
One should consider the whole shape of potential energy surface including anharmonicity of atomic vibrations.

\begin{acknowledgments}
We thank Lucy D. Whalley and Ji-Sang Park for valuable discussions, and Audrius Alkauskas for helpful comments on our manuscript.
This work was supported by the EU Horizon2020 Framework (STARCELL, Grant No. 720907).
Additional funds were received from the Creative Materials Discovery Program through the National Research Foundation of Korea (NRF) funded by Ministry of Science and ICT (2018M3D1A1058536).
Via our membership of the UK's HEC Materials Chemistry Consortium, which is funded by EPSRC (EP/L000202), this work used the ARCHER UK National Supercomputing Service (http://www.archer.ac.uk).
\end{acknowledgments}

\bibliography{bib}

\end{document}